\documentclass[prl,twocolumn,showpacs,debugon,nofootinbib]{revtex4}
\usepackage{epsfig}
\newcommand{\bk}{{\bf k}}
\newcommand{\be}{\begin{equation}}
\newcommand{\ee}{\end{equation}}
\renewcommand{\r}{{\bf r}}

\begin{document}


\title{Photon localization and Dicke superradiance in atomic gases 
}
\author{E. Akkermans$^{1,2}$, A. Gero$^{1}$ and R. Kaiser$^{3}$}
\affiliation{$^{1}$Department of Physics, Technion Israel Institute of Technology,
  32000 Haifa, Israel \\ $^2$Department of Applied Physics and Physics, Yale University, USA   \\ $^{3}$Institut Non Lin\'{e}aire de Nice, UMR 6618 CNRS, France}

\begin{abstract}
Photon propagation in a gas of $N$ atoms is studied using an effective Hamiltonian describing photon mediated atomic dipolar interactions. The density $P(\Gamma)$ of photon escape rates is determined from the spectrum of the $N \times N$ random matrix $\Gamma_{ij} = \sin (x_{ij}) / x_{ij}$, where $x_{ij}$ is the dimensionless random distance between any two  atoms.  Varying disorder and system size, a scaling behavior is observed for the escape rates. It is  explained using microscopic  calculations and a stochastic model which emphasizes the role of cooperative effects in  photon localization and provides an interesting relation with statistical properties of "small world networks".  
\end{abstract}


\date{\today}

\maketitle


We address the question of photon localization resulting both from disorder and cooperative effects  (multi-atomic coherent emission) in three-dimensional cold atomic gases. This localization shows up as an overall decrease of photon escape rates and our purpose is to investigate the roles played by disorder and by cooperative effects. For weak disorder, incoherent spontaneous emission by independent atoms is expected. For stronger disorder, cooperative effects become important and lead to vanishing  escape rates, {\it i.e.} to photons trapped in the gas for very long times. We show that localization occurs as a smooth crossover between these two limits and not as a phase transition like for Anderson localization \cite{anderson}. We propose a stochastic description of the photon emission process, which agrees quantitatively with our numerical results and explains the nature of the crossover. This leads to the conclusion that photon localization in atomic gases is primarily determined by cooperative effects and not by disorder.  

We consider a collection of $N$ identical atoms at rest, taken to be degenerate two-level systems respectively denoted for the atom $i$, by $|g_i \rangle= |j_{g}=0,m_{g}=0 \rangle$ and $|e_i
\rangle = |j_{e}=1,m_{e}=0,\pm1 \rangle$ for  the ground and excited states. $j$ is the total angular momentum and $m$ is its  projection
on a quantization axis. The energy
separation between the two levels, including the radiative shift,  is
$\hbar\omega_{0}$ and the natural width of the excited level is
$\hbar \Gamma_0$. Atoms randomly placed at positions $\r_i$, are coupled to the electromagnetic field ${\bf E}$ through their dipole operator ${\bf d}_i$. The corresponding Hamiltonian is,
\be
H = \sum_{i=1}^N \hbar \omega_0 |e_i\rangle \langle e_i | + \sum_{\bk \varepsilon} \hbar  \omega_k  a_{\bk \varepsilon}^{\dagger} a_{\bk \varepsilon}  - \sum_{i=1}^N {\bf d}_i \cdot {\bf E}(\r_i) 
\ee
where $a_{\bk \varepsilon} ^\dagger$ is the creation operator of a photon of wave vector $\bk$, $ \omega_k = c |\bk |$ and polarization $\varepsilon$. We assume that only one photon is present. For resonant scattering, tracing over the photon degrees of freedom leads to the effective atomic Hamiltonian :
\be
H_{e} = \left(\hbar \omega_0 - i  {\hbar \Gamma_0 \over 2} \right) S_z  + { \hbar \Gamma_0 \over 2} \sum_{i\neq j}  V_{ij} S_i^+ S_{j} ^- 
\label{hamilteff}
\ee
where $S_i^{+} = |e_i \rangle \langle g_i |$ is the atomic raising operator, $S_i^- = (S_i^+ )^\dagger$ and $S_z = \sum_{i=1}^N S_{zi}$ with $S_{zi} =  |e_i \rangle \langle e_i | $. The potential $V_{ij} =  \beta_{ij} - i \gamma_{ij}$ is a random and complex valued quantity, specified by
\begin{eqnarray}
\beta_{ij} &=& {3 \over 2} \left[ -p {\cos k_0 r_{ij} \over k_0 r_{ij}} + q \left(  {\cos k_0 r_{ij} \over (k_0 r_{ij})^3} + {\sin k_0 r_{ij} \over (k_0 r_{ij})^2} \right) \right] \nonumber \\
\gamma_{ij} &=&   {3 \over 2} \left[ p {\sin k_0 r_{ij} \over k_0 r_{ij}} -  q \left( {\sin k_0 r_{ij} \over (k_0 r_{ij})^3} - {\cos k_0 r_{ij} \over (k_0 r_{ij})^2} \right) \right]
\label{deltagamma}
\end{eqnarray}
where $k_0 = \omega_0 / c$ and $r_{ij} = | \r_i - \r_j|$ is the distance between any two atoms. The quantities $p$ and $q$ depend on the atomic transition. For $\Delta m=m_e-m_g=0$, 
$
p_{0} = \sin^{2} \theta_{ij}$  and  $q_{0} = 1 - 3 \cos^2\theta_{ij} $ and for $\Delta m=\pm1$, $p_{\pm} = \frac{1}{2}(1+\cos^2\theta_{ij}) , 
q_{\pm} = \frac{1}{2}(3\cos^2\theta_{ij} -1)$, where $\theta_{ij} = \cos^{-1} (\hat z \cdot \hat{\r}_{ij} )$ and $\hat{\r}_{ij}$ is a unit vector defined along the direction joining the two atoms. Expressions (\ref{hamilteff}) and (\ref{deltagamma}) are well known \cite{lemberg} and correspond to an  instantaneous photon exchange between an initially excited atom $i$ coupled to the $j$-{\it th} atom. This description holds for distances between atoms that are small compared to the coherence length of the light emitted by a single atom \cite{miloni}.

The possible escape rates $\Gamma$ of a photon propagating in an atomic gas, are obtained from the time evolution of the ground state population $\langle G | \rho | G \rangle$,
\be
{ d \over dt} \langle G | \rho | G \rangle = \Gamma_0 \sum_{ij} \gamma_{ij} \langle G | S_j ^- \rho S_i ^+ | G \rangle
\label{popt}
\ee
where $|G \rangle \equiv | g_1, g_2, \dots , g_N \rangle$ and 
$\rho (t)$ is the reduced atomic density operator \cite{stehle}. The real and symmetric matrix $\gamma_{ij}$ can be diagonalized by an orthogonal transformation. The eigenvalue equation 
\be
\sum_{j=1}^N \gamma_{ij} u_j ^{(k)} = \Gamma_k  u_i ^{(k)} \, \ 
\label{eigen}
\ee
together with the collective operators $S_k ^\pm \equiv \sum_{i=1}^N u_i ^{(k)} S_i ^\pm$, enable to rewrite (\ref{popt}) as $\langle G | \dot{\rho} | G \rangle = \Gamma_0 \sum_{k=1}^N \Gamma_k  \langle G | S_k ^- \rho S_k ^+ | G \rangle$. This last form allows to interpret the eigenvalues $\{ \Gamma_k \}$ as  the escape rates and the eigenfunctions $\{ u_i ^{(k)} \}$ as photon modes providing the directivity of the angular emission \cite{stehle}. It results from (\ref{popt}) that the $ \Gamma_k $'s are independent of the effective dipole-dipole interaction $\beta_{ij}$  so that Van der Waals dephasing does not play a role \cite{haroche}.

The average density of escape rates $P (\Gamma)$, normalized to unity, is defined by $P(\Gamma) = -(1/ \pi) \, \mbox{Im} R(z = \Gamma + i 0^+ )$, with 
\be
R(z) = {1 \over N} \overline{\mbox{Tr} \left( {1 \over z - {[ \gamma_{ij} ]}} \right)} \, \ .
\label{resolvent}
\ee
The average $\overline{\cdots}$ is taken, at fixed density, over spatial configurations of the atoms. For Gaussian ensembles of random matrices \cite{mehta}, $P(\Gamma)$ obeys a semi-circle law. Here, as we shall see, the behavior is very different. 

The random matrix $\gamma_{ij}$ depends on the distances $r_{ij}$ between atoms and on the angles $\theta_{ij}$.  We expect localization properties to depend on $r_{ij}$ rather than on $\theta_{ij}$. We therefore consider  the scalar model \cite{eaag} obtained from (\ref{deltagamma}) by averaging $\gamma_{ij}$ over $\theta_{ij}$ and thus given by the $N \times N$ random matrix  \cite{mezard},
\be
\Gamma_{ij} \equiv \langle \gamma_{ij}  \rangle = {\sin x_{ij} \over x_{ij}} 
\label{scalar}
\ee
where $x_{ij} = k_0 r_{ij}$ are interatomic distances expressed in units of the wavelength $\lambda = 2 \pi / k_0$. We have checked that the vectorial  and scalar behaviors of $P(\Gamma)$ are  qualitatively the same \cite{tbp}. This point is important since one may think that the scalar case (\ref{scalar}) is the far-field limit obtained by dropping in (\ref{deltagamma}) the near-field terms responsible for  cooperative effects. This is not the case and $\Gamma_{ij}$ is well defined for $x_{ij} =0$ so that $\mbox{Tr} [ \Gamma_{ij} ] = N$.  The eigenvalues of $\Gamma_{ij}$ are nonnegative since the $3d$ Fourier transform of $\mbox{sinc} |x|$ is $\delta (|k| -1) \geq 0$. 

We now consider $N$ atoms enclosed in a cubic volume $L^3$, with a uniform density $n$.  The disorder strength is defined by the dimensionless parameter $W = 1 / k_0 l$ where $l = 1/ n \sigma$ is the elastic mean free path and $\sigma \simeq \lambda^2 $ is the resonant scattering cross section \cite{am}. Introducing the number $N_\perp \equiv (k_0  L)^2 /4$ of transverse photon modes, leads to $W =  {\pi \over 2} {\lambda \over L} {N \over N_\perp}$. 

\begin{figure}[ht]
\centerline{ \epsfxsize 9.5cm \epsffile{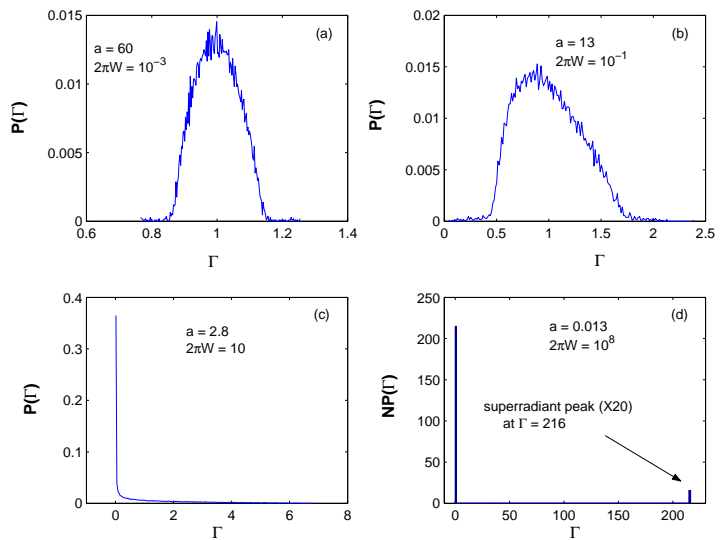} } \caption{\em
Behavior of $P (\Gamma)$ for different values of disorder $W$ and size $a = L / \lambda$, with $N = 216$. (a) low disorder (b), (c) larger disorder and (d) Dicke limit. }
\label{Eigenvalues}
\end{figure}

 Characteristic behaviors of $P(\Gamma)$ for different values of $W$ and size $a= L / \lambda$ are displayed in Fig.\ref{Eigenvalues}. For a dilute gas ($W \ll 1$) we recover the single atom limit $ \Gamma_{ij} \rightarrow \delta_{ij}$, namely $P(\Gamma)$ is narrowly peaked around $\Gamma = 1$ (in units of $\Gamma_0 $) as expected from resonant scattering of a photon by a single atom (Fig.\ref{Eigenvalues}.a). For stronger disorder, $P(\Gamma)$ becomes broader and shifted towards lower values of $\Gamma$ (Fig.\ref{Eigenvalues}.b). Eventually for large enough disorder, most of the eigenvalues get close to $\Gamma =0$ (Fig.\ref{Eigenvalues}.c). Such a vanishing escape rate corresponds to photons localized in the atomic gas. By increasing further $W$, at fixed number $N$ of atoms, yet another behavior shows up for $x_{ij} \ll 1$ (Fig.\ref{Eigenvalues}.d) where $P(\Gamma)$, obtained from the escape matrix with all entries equal to one ($ \Gamma_{ij} = 1$), has two eigenvalues. One at $\Gamma =0$ is the $(N-1)$-degenerate subradiant mode and the second $\Gamma = N$ is the non-degenerate superradiant mode.  This is the Dicke limit \cite{dicke} reached when atoms are enclosed in a volume much smaller than $\lambda^3$. Using (\ref{resolvent}), we obtain the density:
\be
P(\Gamma) = {N-1 \over N} \delta (\Gamma) + {1 \over N} \delta (\Gamma - N) \, \ .
\label{dicke}
\ee

\begin{figure}[ht]
\centerline{ \epsfxsize 7cm \epsffile{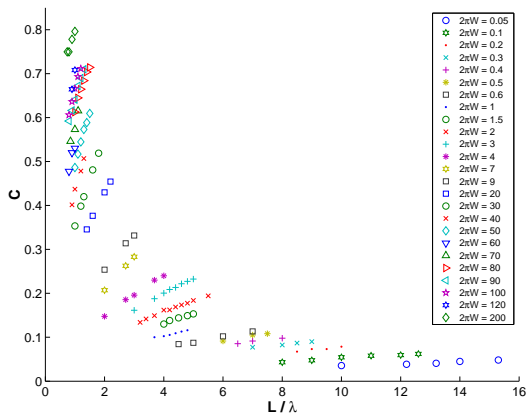} } \caption{\em Behavior of $C$ as a function of the size $a = L / \lambda \geq 1$, for different disorder strengths $W$. 
 }
\label{Scaling1}
\end{figure}

\begin{figure}[ht]
\centerline{ \epsfxsize 7.5cm \epsffile{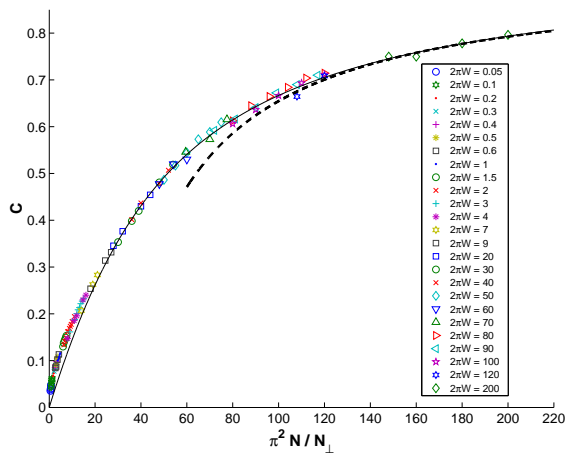} } \caption{\em All points represented in Fig.\ref{Scaling1} collapse on a single curve as expected from (\ref{scaling}). The solid line is a fit of (\ref{sw}) and the dashed line is the asymptotic behavior obtained from (\ref{asympt}).
 }
\label{Scaling2}
\end{figure}

\begin{figure}[ht]
\centerline{ \epsfxsize 6cm \epsffile{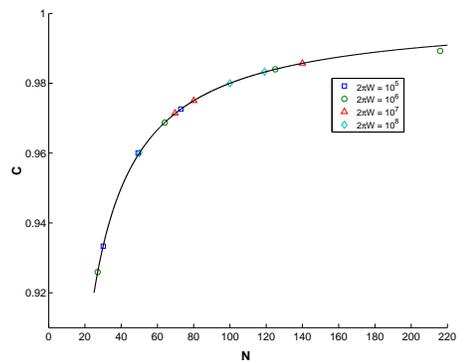} } \caption{\em Behavior of $C(N)$ in the Dicke limit. 
 }
\label{Dicke}
\end{figure}

A quantitative characterization of the behavior of $P(\Gamma)$ is obtained by considering the relative number of states 
$ \int_1 ^\infty d \Gamma \, \ P(\Gamma)$ having an escape rate larger than 1. We then introduce the conveniently normalized function $C(a,W)$ defined between 0 and 1 by
\be
C(a,W) = 1  -  2  \int_1 ^\infty d \Gamma \, \ P(\Gamma) \, \ .
\label{g}
\ee
$C(a,W)$ thus defined, measures the relative number of states having  a vanishing escape rate. At finite size, we expect $C(a,W)$ to have a scaling form \cite{rqcutoff}, namely : 
\be
{d \ln C(a,W) \over d \ln a} = \beta (C)
\label{scaling}
\ee
whose solution $C(a,W)$ is a function of $a/ \xi(W)$ alone. We have verified this scaling behavior over a broad range of size and disorder. For $a \geq 1$, the results displayed in Fig.\ref{Scaling1}, collapse on a single curve (Fig.\ref{Scaling2}) when plotted as a function of the  parameter $2 \pi aW =  \pi^2  N / N_\perp$.  
The Dicke limit (Fig.\ref{Eigenvalues}.d) is reached for small volumes, $L \ll \lambda $, and $N_\perp \simeq 1$, so that the scaling parameter becomes $N / N_\perp \simeq N$. Using (\ref{dicke}) and (\ref{g}), leads to $C(a,W) = 1- (2 /N) $ displayed in Fig.\ref{Dicke}.  

To explain this scaling behavior, we study first  the limit $N \leq N_\perp$ which can be understood from simple considerations on the superradiant rate. The electric field $E$ created by $N$ excited atoms within the volume $L^3$, is obtained from the electromagnetic energy $N \hbar \omega_0 = L^2 (c / \Gamma) E^2 / 8 \pi$. The superradiant escape rate $\Gamma$ may also be obtained by assuming that under the action of the field $E$, each atom performs half a Rabi oscillation over a  time $1/ \Gamma = \hbar /E d$. The dipolar matrix element $d $ is related to the spontaneous emission rate $\Gamma_0 = {4 \over 3} {k_0 ^3 d^2 \over \hbar}$. Altogether, $\Gamma$ is specified by the parameter $ { N \over N_\perp}$ as
\be
{\Gamma \over  \Gamma_0  } = {3 \pi \over 2} { N \over N_\perp}  \, \ .
\label{smallvalue}
\ee
This expression which is correct when the electric field $E$ is delocalized over the atoms {\it i.e.} for $N / N_\perp \leq 1$, emphasizes that the initial linear behavior of  $C (N / N_\perp)$ (Fig.\ref{Scaling2}) is essentially determined by cooperative effects. In the opposite limit $N \gg N_\perp$, the $n$-th order cumulant of $P(\Gamma)$, is $(1/N) \overline{ \mbox{Tr} \, \Gamma_{ij} ^n } = 3 (N / N_\perp)^{n-1} / (n+2)$, and a resummation leads to the asymptotic behavior \cite{tbp}, 
\be
P(\Gamma) =  ( 1 - {3 N_\perp \over 2 N}) \delta(\Gamma) + 3 \Gamma ( { N_\perp \over N} )^3 \label{asympt} \ee for $ \Gamma \leq {N / N_\perp}$ and $P(\Gamma) = 0$ otherwise, so that asymptotically, $C(N/N_\perp) = 1 - 3 (N_\perp / N)$ (Fig.\ref{Scaling2}). 

To interpolate  between the two previous limits, we consider a mapping of the cooperative emission of randomly distributed atoms onto a stochastic Markov process on a graph. To define it, we start from the  Dicke limit \cite{dicke} whose escape rate matrix $\Gamma_{ij} = 1$, is the adjacency matrix of a complete graph having spectral density (\ref{dicke}). In that  limit, the Hamiltonian (\ref{hamilteff}) commutes with the collective spin operators $S^\pm \equiv \sum_{i=1}^N S_i ^\pm$, so that Dicke states $| S,m \rangle$, eigenstates of $S^2 \equiv {1 \over 2} ( S^+ S^- + S^- S^+ ) + {1 \over 4} S_z ^2$ and $S_z $, are eigenstates of $H_e$. The emission of a photon, a process which keeps $S$ unchanged and shifts $m$ by one unit, can then be described as a one-dimensional Markov process, $P_N (0)$, with equiprobable jumps $(1/2)$ between neighboring $m$-states.  Away from the Dicke limit, the expressions of the collective spin operators for scalar waves, are  obtained by taking $u_i ^{(k)} = e^{\pm i \bk \cdot \r_i} $ in (\ref{eigen}), so that $S_{\bk} ^\pm = \sum_{i=1}^N S_i ^\pm e^{\pm i \bk \cdot \r_i} $.  The random phases prevent  from having the previous angular momentum algebra so that $H_e$ does not commute with $S_{\bk} ^2$. Yet, we can still use the Dicke states basis, but now a  photon emission is a process where both $S$ and $m$ change. This can be described as a modified Markov process where $S$-changing events are accounted by adding random jumps to non-neighboring $m$-states with a  probability $\epsilon$. The corresponding Markov process $P_N (\epsilon)$ is defined by the $N \times N$ matrix 
\be
P_N (\epsilon ) = (1 - N \epsilon) P_N (0) + \epsilon |e_N \rangle \langle e_N |
\label{markov}
\ee
where $P_N (0)_{i, i\pm1} = 1/2$. The escape rate of a photon is set by the inverse mean hitting time $ {\cal T}_N (\epsilon)$ defined as the average over all sites of the number of steps needed to reach a given assigned  site \cite{simone}. The scaling function (\ref{g}) can be written as $C(\epsilon, N) = \epsilon N  { \cal T}_N ^2 (\epsilon) / { \cal T}_N (0)$ and for large $N$, $C$ depends on  $s \equiv (\gamma / \tanh \gamma) -1$ alone, namely 
\be
C(s) =  {s^2 \over \gamma^2 (s)} 
\label{sw}
\ee
where $\gamma \equiv \sqrt{\epsilon N^3 / 2}$ \cite{tbp,simone}. The parameter $s$ is the number of shortcuts between non neighboring $m$-states induced by emission processes which do not leave $S$ unchanged. A plot of (\ref{sw}) is shown in Fig.\ref{Scaling2} and compared to the numerical data.

\begin{figure}[ht]
\centerline{ \epsfxsize 5.2cm \epsffile{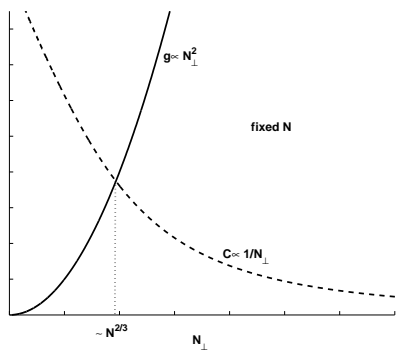} } \caption{\em Comparison between cooperative effects determined by the parameter $N / N_\perp$ and disorder described by $g = N_\perp ^2 / N$. 
 }
\label{disorder}
\end{figure}

Previous expressions indicate that photon escape rates are primarily  determined by cooperative effects and not by disorder. Moreover, (\ref{sw}) displays a smooth crossover between delocalized and localized photons and not a disorder driven localization transition. To emphasize this point, we compare the two parameters, $N / N_\perp$ which specifies the strength of cooperative effects, and the dimensionless conductance $g \equiv N_\perp ^2 / N$ which determines  the strength of disorder \cite{am}. These two quantities are plotted in Fig.\ref{disorder} for a fixed, but large $N$. For $N_\perp \geq N^{2/3}$, $g$ is large so that disorder effects are small. In the opposite limit, we have $N / N_\perp \gg g$ so that cooperative effects prevail disorder in a regime ($g \ll 1$) where disorder may precisely be strong enough to lead to disorder driven localization. 

To summarize, we have shown that escape rates of photons propagating in a $3d$ atomic gas, are characterized by a scaling behavior determined by the parameter $N/ N_\perp$. For $N / N_\perp \ll 1$, cooperative effects are negligible and photons are emitted in spontaneous and incoherent processes. For larger values of this parameter, cooperative effects set in and the overall photon escape rate becomes very small.  Eventually, for fixed $N$, we reach the Dicke limit.  The crossover between these behaviors is smooth and a disorder induced localization phase transition is unlikely to take place.

These results are well described by the modified Markov process (\ref{markov}). Surprisingly enough, this process also provides an accurate description and an interesting link \cite{simone} to the recently widely studied "small world networks" \cite{barabasi}, tuned to be intermediate between regular networks with long ($\propto N$) chemical lengths and random networks with short ($\propto \ln N$) ones. The existence of a crossover rather than a phase transition, between regular and random networks has been thoroughly investigated \cite{moore}. Noting that "small world networks" appear to be well suited to achieve synchronization of non linear oscillators \cite{strogatz} and that photon cooperative emission results from the synchronization of the atomic dipoles induced by long range atomic correlations, this connexion becomes even more interesting and relevant. Finally, the analysis we present in this letter may suggest a different approach and new protocoles for experiments on photon localization in cold atomic gases.

This research is supported in part by the Israel Academy of
Sciences, by the Fund for Promotion of Research at the
Technion and the ANR program CAROL (ANR-06-BLAN-0096).


\begin{thebibliography}{99}
\bibitem{rqcutoff} Scaling has been shown to be independent of the choice of the cutoff in the lower bound of the integral in (\ref{g}) \cite{tbp}.
\end{thebibliography}
    \end{document}